\documentclass[journal]{IEEEtran}
\IEEEoverridecommandlockouts
\usepackage[utf8]{inputenc}
\usepackage{cite,balance}
\usepackage{amsmath,amssymb,amsfonts,multirow,upgreek,subcaption}
\usepackage[table]{xcolor}
\usepackage{textcomp}
\usepackage{xcolor}
\usepackage{bm}
\usepackage{algpseudocode}
\usepackage{algorithm}
\usepackage{enumitem}
\usepackage{psfrag}
\usepackage{tikz}
\usetikzlibrary{shapes.geometric, arrows}
\tikzstyle{roundrec} = [font=\footnotesize,rectangle, rounded corners, minimum width=1cm, minimum height=1cm,text centered, text width=2.5cm, draw=black]
\tikzstyle{rec} = [font=\footnotesize,rectangle, minimum width=3cm, minimum height=1cm, text width=3.75cm, draw=black]
\tikzstyle{bubble} = [font=\footnotesize,rectangle, rounded corners, text centered, draw=black]
\tikzstyle{arrow} = [thick,->,>=stealth]
\def\BibTeX{{\rm B\kern-.05em{\sc i\kern-.025em b}\kern-.08em
    T\kern-.1667em\lower.7ex\hbox{E}\kern-.125emX}}

\title{Extended Automatic Repeat Request For Integrated Sensing And Communication Networks}

\author{Georgios Mylonopoulos, Behrooz Makki, \emph{Senior Member IEEE}, Stefano Buzzi, \emph{Senior Member IEEE}, G\'abor Fodor, \emph{Senior Member IEEE} \vspace{-1cm}
\thanks{This work has been supported by the EU Horizon 2020
MSCA-ITN-METAWIRELESS, Grant Agreement 956256. 
}
}
\date{January 2024}

\begin{document}

\maketitle

\begin{abstract}
6G wireless networks will integrate communication, computing, localization, and sensing capabilities while meeting the needs
of high reliability and trustworthiness.
In this paper, we develop similar techniques as those used by communication modules of previous generations 
for the sensing functionality of 6G networks. 
Specifically, this paper introduces the concept of extended automatic repeat request (e-ARQ) 
for integrated sensing and communications (ISAC) networks. 
We focus on multi-static sensing schemes, in which the nodes receiving the reflected sensing signals provide 
the transmitting nodes with configurable levels of feedback about the sensing result. 
This technique improves the sensing quality via retransmissions using adaptive parameters. 
We show that our proposed e-ARQ boosts the sensing quality in terms of detection accuracy 
and provides a sense of adaptability for applications supported by ISAC networks.
\end{abstract}

\section{Introduction}
For decades, wireless communication and radar technologies have coexisted with limited mutual impacts. 
The main regulatory and standardization efforts have been on interference avoidance and management via means, such as allocating separate frequency bands to these two technologies, such that they can coexist without causing interference to each other.
With the arrival of 5G and beyond technologies, and due to recent advances in radar systems, the operational frequencies of these systems are merging; key radar bands may reuse some of the millimeter wave (mmWave) wireless communication bands. 
For example, the radar bands K (18 GHz-26.5GHz) and Ka (26.5 GHz – 40 GHz) are close to mmWave wireless communication bands. 
Moreover, with the probable use of sub-THz bands in 6G, it will be possible to perform accurate sensing using cellular base stations and wireless access points. In particular, 6G systems will use a massive number of antennas and wide range of spectrum, which will enable to deploy integrated sensing and communication (ISAC) systems and offer accurate sensing services in the wireless network~\cite{Wei:22, Wang:23}.

With ISAC, the objective is to share the spectrum between sensing and communication and reuse the existing wireless network infrastructure for sensing functionalities. In other words, ISAC refers to adding sensing capabilities and radar-like functionalities to wireless networks and  thereby creating perceptive wireless and cellular networks capable of
detecting and tracking connected and unconnected objects (targets) within the service area of the network~\cite{guo2024integrated}. 

A deep overview of ISAC networks can be found in, for example,~\cite{zhang2021overview,zhang2021enabling,liu2022integrated}. 
Also, initial testbed evaluations have been presented in, e.g.,~\cite{zhang2021design} and~\cite{colpaert2023massive}, where the integration not only boosts the sensing quality, but also improves the communication performance. 
ISAC presents both opportunities and challenges. The main opportunity is that the sensing and communication functionalities can be performed in the same nodes, which enables centralized resource allocation and interference management under network control. 
Also, compared to the deployment of a separate network for sensing, the integration of the sensing capability can be introduced at a low cost by reusing the existing wireless network infrastructure. 

One of the challenges of the ISAC is that, different from sensing, communication is mainly based on standardized operations. Therefore, the convergence requires extensive standardization efforts to include the sensing. 
For instance, 3GPP has recently started a study-item on ISAC in Release 19. Other challenges are interference management and resource allocation. In particular, ISAC must deal with time, frequency, and/or energy trade-offs, where limited resources should be properly shared between each of the sensing and communication functionalities. Finally, depending on the deployment, (self-)interference due to full-duplex operation may be challenging, although proper beamforming and isolation can minimize the full-duplex effects~\cite{barneto2021full,Smida:23}.

In general, sensing methods can be divided into two categories. 
With mono-static sensing, the transmission of the sensing signal and the reception of the reflected signal 
are performed by the same node. 
With multi-static sensing, on the other hand, the transmission and the reception are performed by different nodes. 
The simplest type of multi-static sensing is bi-static sensing with only two nodes being involved in the sensing process. Compared to mono-static sensing, multi-static sensing is less sensitive to self-interference, 
at the cost of tight coordination and synchronization between the nodes.
For these reasons, multi-static sensing is of more interest in practice. 

One of the benefits of ISAC is that the wireless communication techniques, which have been already developed and optimized over the decades, can be extended for sensing functionalities. With this background, the present paper introduces extended version of automatic repeat request (e-ARQ) process for sensing functionalities. With typical ARQ in wireless networks, depending on the message decoding status, the receiving (Rx) node sends acknowledgement/negative acknowledgement (ACK/NACK) to the transmitting (Tx) node and, with NACKs, the retransmissions continue until the message is correctly decoded or the maximum number of retransmissions is reached. There are different types of ARQ where either the same or new redundancy bits for the failed signal are retransmitted, and the receiver adapts the decoding method based on the considered type of ARQ (see, e.g.,~\cite{caire2001throughput,makki2014performance} for an overview of ARQ protocols). As opposed to communication networks, with sensing there is no data to be decoded by the Rx node. As a result, unlike wireless networks, the feedback to the transmitter may contain different information about the sensing quality and/or the ability/disability of the Rx node to accurately sense the targeted objects.

Moreover, as explained in the following, because there is no information to be decoded by the sensing Rx node, the adaptation protocols in the retransmissions of the sensing signal are inherently different from the ones adapted for typical ARQ protocols in communication networks. As we show, handshaking between the Tx and Rx nodes of the sensing signal about the sensing result makes it possible to adapt the transmission and reception/sensing configurations at the Tx and Rx nodes, respectively, which improves the sensing quality significantly. This, in turn, improves the network spatial understanding which, concurrently, results in better communication performance.

\section{E-ARQ for Sensing}
For simplicity, we explain the setup for the cases with bi-static sensing method with a Tx node sending a sensing signal to be reflected by the target and received by the Rx node. However, our proposed method is also well applicable to the cases with multi-static sensing.

With a multi-static or bi-static sensing, it is probable that the Rx node may not be able to sense the target properly, does not sense a target or loses the track of the target. In such cases, it is beneficial to follow the same procedure as in the ARQ methods used in wireless communications to inform the Tx node, such that the sensing signal is retransmitted with proper reconfigurations. Also, with multiple (re)transmissions, the Rx node can benefit from the combination of the signals received in different retransmission rounds, and improve the sensing quality by applying a proper sensing method combining different received signals.

\begin{figure}
\centering
\includegraphics[width=0.475\textwidth]{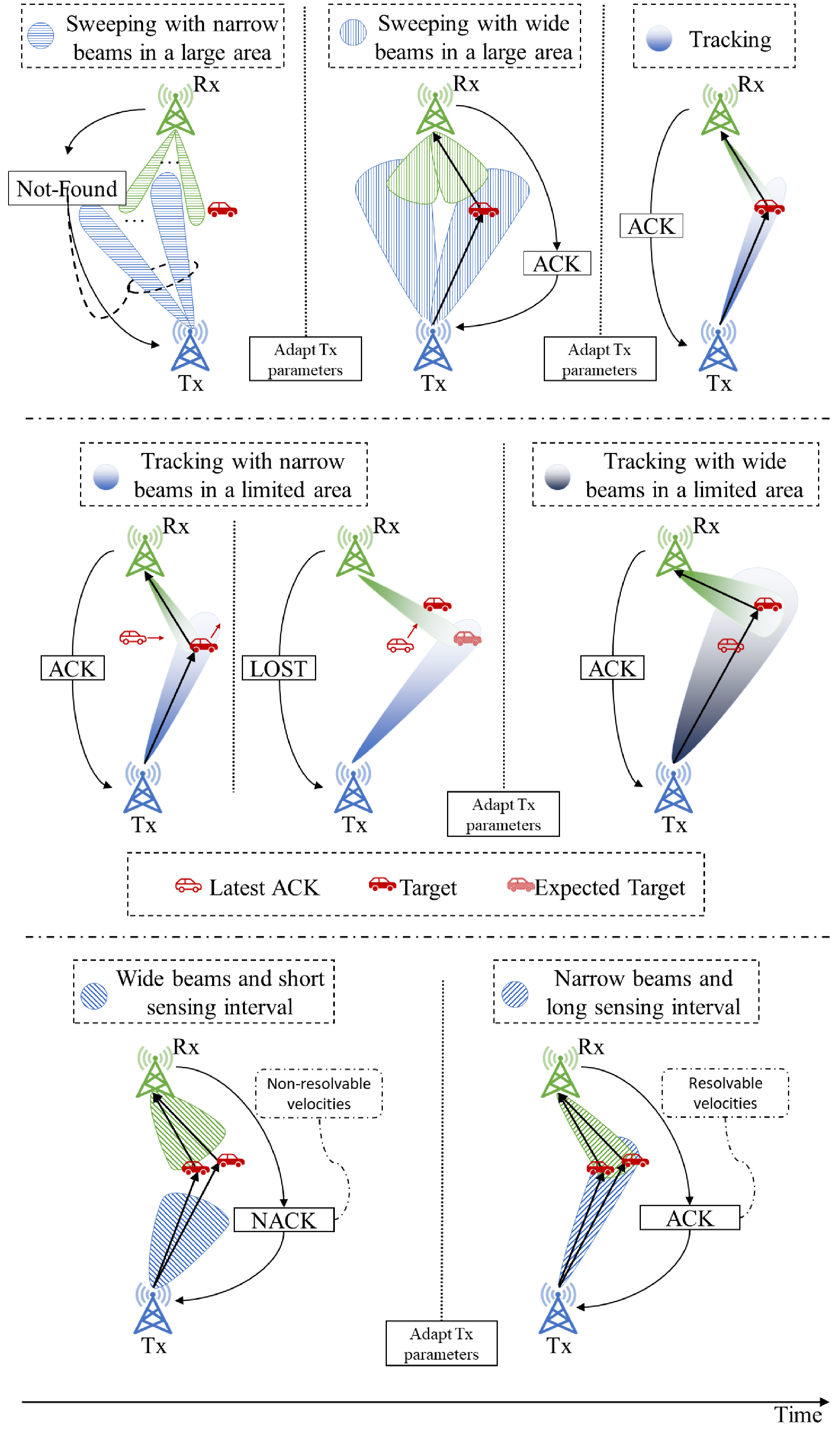}
\caption{Different use-case scenarios for different bi-static sensing applications where an e-ARQ procedure is implemented. The figures should be read from left to right. The e-ARQ procedure triggers sensing adjustments, when necessary, for the Tx and Rx nodes and validates different node configurations that lead to proper sensing.}
\label{fig:usecase}
\end{figure}

Figure~\ref{fig:usecase} illustrates our proposed e-ARQ concept. With e-ARQ for sensing, once the Rx node has received a sensing signal (or, has received no signal in resource blocks it expects to receive the signal), it sends feedback to the Tx node about the sensing result. By sensing result, one may consider different metrics such as the positioning accuracy, the positioning resolution, the velocity accuracy, the velocity resolution, etc., or their combination. As demonstrated in Fig. 1, depending on the sensing result, different feedback may be sent to the Tx node: 
\begin{enumerate}
    \item[a.] ACK: Used when the target is sensed with high quality and, thereby, no more retransmissions are required.
    \item[b.] NACK: Used when the target is sensed but the sensing quality is low/unreliable and, thereby, retransmissions and/or some sensing configuration adjustments may be required.
    \item[c.] LOST: Used when an already sensed target is lost, e.g., during the tracking process and, thereby, retransmissions and/or drastic sensing configuration adjustments may be required.
    \item[d.] Not-found: Used when no target is sensed or no sensing signal is received in the resource blocks that the Rx node expects to receive the signal.
\end{enumerate}

In the first example of Fig.~\ref{fig:usecase}, the Rx node initially receives no sensing signal in the expected resource blocks (via sensing scheduling) and, thereby, informs the Tx node that no target is sensed via a Not-found signal. As a result, the nodes may adapt the transmission configurations, e.g., the Tx/Rx beams are widened, and retransmit the sensing signal. In the second example, an already sensed target may get lost during the tracking process due to, e.g., blockage, unpredicted trajectory, etc. As a result, the Rx node feedback LOST to indicate the need for re-calibrating the tracking parameters. Then, for instance, the Tx node may adjust the target’s trajectory prediction and utilize wider beams for retransmissions. Finally, in the third example, the Rx node may not be able to resolve the sensed targets’ velocities and, therefore, informs the Tx node via a NACK response, that, e.g., longer sensing interval is required and narrower -better focused- beams. Accordingly, the transmission and/or reception/sensing methods may be adapted at the Tx and Rx nodes, respectively. The retransmissions continue until the target is properly sensed, the maximum number of retransmissions is reached, or the target is estimated to have moved out of the sensing coverage area of the gNB.

Depending on the received feedback, one may consider different methods for reconfiguration of the retransmission signals. For instance, with a Not-found, the retransmissions may be based on wider Tx/Rx beams, and the beams may cover different directions. As opposed, with a NACK, the narrow beams may be used by the Tx node focusing on specific direction, because the rough position of the target is already known and the objective is to improve the sensing accuracy. Finally, with a LOST, the Tx node may use wide or semi-wide beams to cover specific areas predicted from the moving trajectory of the target before it was lost. Also, following the same methods as in typical ARQ schemes in wireless network~\cite{makki2013green}, with LOST, Not-found or NACK, the Tx node may increase the sensing transmit power gradually in the retransmissions. Alternatively, different (random or predefined) polarizations, beam directions or beam widths may be used in the retransmissions to generate diversity. On the other hand, with an ACK, the Tx node may narrow down the beams in specific area or reduce the Tx power, etc.

During the retransmissions, the parameters of the sensing signal, e.g., the duration, the bandwidth, the periodicity of the sensing signal, the Tx power, may be adapted (see~\cite{behravan2022introducing} for the appropriate sensing signal design for different quality-of-service requirements). Particularly, depending on the considered unambiguous range, range resolution, unambiguous velocity and/or velocity resolution, the transmissions may start with aggressive parameter settings and continue conservatively during the retransmissions.

The sensing method may be adapted by the Rx node during the retransmissions rounds. In its simplest form, the Rx node may retry sensing by considering only the latest received sensing signal and ignoring the previously received signals. This is of low complexity, at the cost of possible multiple retransmissions. Alternatively, for tracking purposes, the Rx node may consider a limited number of latest received sensing signals. Finally, the sensing may be based on the combination of different retransmitted signals where the combinations is used for, e.g., triangulation, accurate line-of-sight (LOS) estimation, etc. This is indeed at the cost of complexity. 

To implement the e-ARQ procedure, different algorithmic approaches may be employed. For instance, dual threshold sequential detection (DTSD) is a sensing technique employed to improve the probability of detection while the false alarm rate is maintained below a desired value, e.g.,~\cite{7888967}. In the traditional DTSD, a test parameter, e.g., the received energy, is evaluated against two different thresholds. There are three potential outcomes for the detection hypothesis, rather than two, i.e., the traditional pass-fail outcomes and a third option indicating a possible detection, when the test parameter is above the fail threshold, but below the pass threshold.
Additionally, we propose an extended, memory-enabled DTSD (ME-DTSD), in which the sensing memory is included into the decision making process, augmenting the number of potential outcomes.

We propose to extend the ME-DTSD scheme into the algorithmic back bone of the e-ARQ procedure (see Section III for an example). The strong pass threshold could produce an ACK signaling, indicating the proper detection/inspection of a target. As discussed above, the concept of proper inspection can be defined to fit the system's requirements and capabilities. Moreover, the possible detection defined by a test parameter being between the two thresholds can lead to a NACK signaling. The NACK signaling alters the Tx node and Rx node configurations. A more compatible sensing configuration shall increase the test parameter and eventually result in an ACK. When the test parameter falls below the lower threshold either a LOST or a Not-found signal may be triggered, depending on the previous detection outcomes available at the Rx node. Essentially, when we jointly consider prior sensing outcomes, we enable a more complex e-ARQ procedure that allows for additional degrees of freedom in the design stage.

There are several degrees of freedom in the design of such an e-ARQ procedure and there are benefits that need to be highlighted as its performance is taken into account. In a dynamic system without the adaptability offered by the e-ARQ procedure, as the targets under inspection move and the sensing conditions change (see Fig.~\ref{fig:usecase}), the Tx/Rx node configurations may need to change dynamically to account for all potential sensing requirements.
An \emph{open-loop} is inefficient as resources are wasted to test different configurations, which may not be required. On the other hand, the extended concept of ACK signaling highlights the compatibility of the current configuration and the wasted resources are minimized. As the current configuration becomes less effective, a NACK is expected to trigger the necessary search for a better suited configuration when necessary. Furthermore, introducing a sense of memory into the system and jointly processing past and current sensing outcomes allows the Rx node to detect the system's sensing limitations. This ability is due to the fact that the Rx node can make deductions about the expected position and/or velocity of a known target based on past sensing outcomes and the abrupt disappearance of the sensed target highlights the inherent inability of the system to properly sense targets in specific locations or speeds, e.g., due to blockages. Moreover, adaptive sensing based on e-ARQ reduces the sensing-to-communication interference as, e.g., unnecessary sensing-based sweepings are avoided. Finally, note that the e-ARQ feedback may be received jointly or separately from the ARQ-based feedback associated with communication functionality.

\section{A Toy-example}
To better illustrate the working principle of an adaptive sensing e-ARQ procedure based on an ME-DTSD, we introduce a toy example and analyse its performance. We consider a bi-static ISAC scenario where a communicating user equipment (UE) is onboard a moving vehicle, i.e., the UE and the target are co-located. This is not a necessary assumption, and is considered due to the limit on the number of words/figures as well as to simplify the discussions. The UE communicates with the transmitting gNBs, while a secondary gNBs is in Rx mode, assisting with the bi-static detection, and tracking of the moving vehicle (target). The simulation parameters are presented in Table~\ref{tab:sim_parameters}.

\begin{table}
\caption{Simulation parameters.}
\label{tab:sim_parameters}
\begin{center}
\begin{tabular}{|c c| c c|}
\hline
gNBs Antennas    & $16$                  & Beams (per gNB)       & $18$                  \\
Frequency        & $24$ GHz              & Antenna Spacing       & $\lambda / 2$         \\
Sampling rate    & $10^{-1}$ sec         & Target velocity       & $2.25\;\rm{ ms}^{-1}$ \\
Tx gNB           & $[0, 0]$ m            & Rx gNB                & $[-40, 40]$ m         \\
Small obstacle   & $0.5\times 0.5$ m$^2$ & Large obstacle        & $5\times 5$ m$^2$     \\
Tx Power (init.) & $1\times 10^{-4}$ W   & Tx Power (incr.)      & $3\times 10^{-4}$ W   \\
Bandwidth        & $15$ KHz              & Noise Figure          & $6$ dB                \\
\hline
\end{tabular}
\end{center}
\end{table}

\begin{figure}
    \centering
    \includegraphics[width=0.475\textwidth]{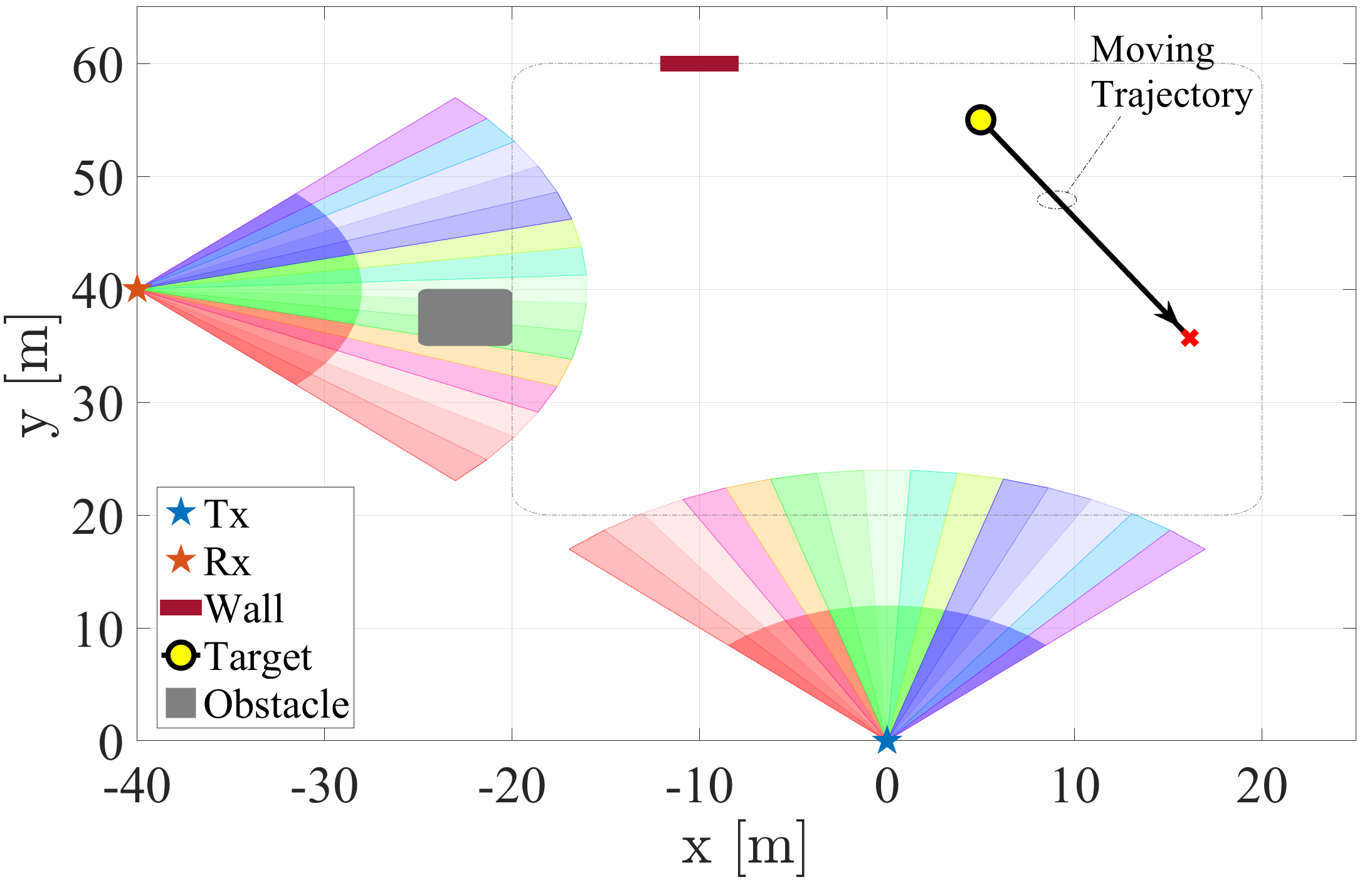}
    \caption{Toy-example geometry. The different coloured angular domains represent the 3-dB beamwidth of different sensing configurations.}
    \label{fig:geometry}
\end{figure}

The 2-dimensional geometry of the considered toy-example is shown in Fig.~\ref{fig:geometry} with the two gNB nodes and the moving UE/target. The Tx and Rx nodes have a number of different beam configurations, targeting different portions of the angular domain. Note that wider beams are also available, with the smaller radius of the corresponding wedge highlighting the smaller gain compared to a well matched narrow beam. The target is initially visible to both the Tx and the Rx nodes. However, as the target moves the Rx node's LOS link is blocked by an obstacle. We present the results for two cases with small and large obstacles, with sizes given in Table~\ref{tab:sim_parameters}. Therefore, the sensing application is expected to be impaired. We also consider a small wall, as shown on top of Fig.~\ref{fig:geometry}, that provides a secondary, unobstructed, link between the target and the Rx node. Both the obstacle's and the wall's presence are not known to the Rx node. 

Intuitively, it is expected that the Tx and the Rx nodes shall require different configurations as the target moves across their corresponding angular domains. From a communication perspective, with each signal transmission we require proper precoding/beamforming at the Tx node, resulting in high signal to noise ratio (SNR) at the UE. On the other hand, from a sensing perspective, with each signal we attempt to capture a reflected target echo, with a corresponding received echo strength indicator (RESI) value, which serves as the test parameter for the detection hypothesis. The RESI is defined as the energy ratio of the received echo to the expected noise. That is, the RESI essentially captures how much energy is received by the Rx node for the Tx-Rx configuration employed at the time.

\begin{figure}
    \centering
    \includegraphics[width=0.475\textwidth]{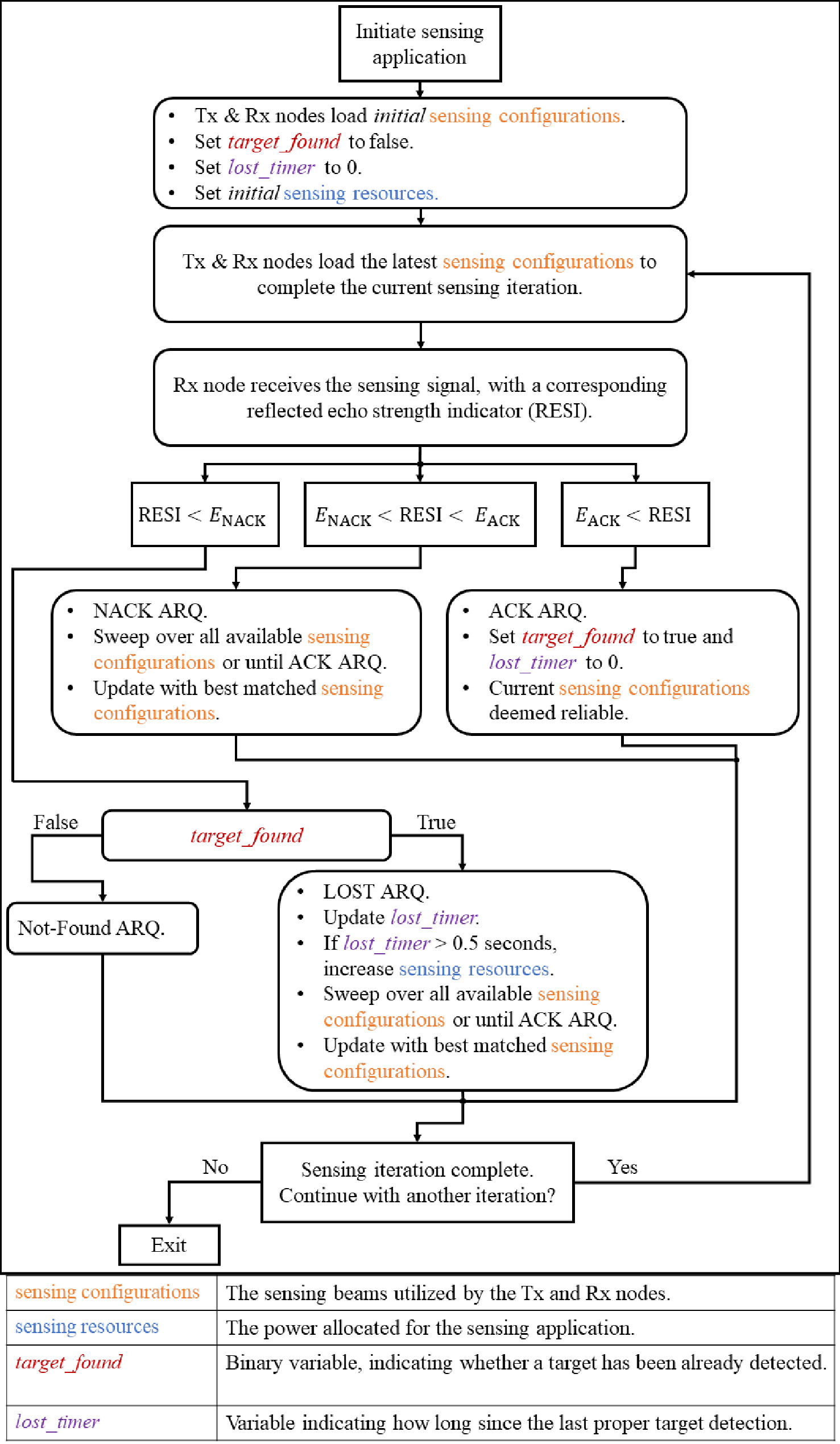}
    \caption{Flowchart for the automated sensing application based on our proposed e-ARQ based sensing.}
    \label{fig:flowchart}
\end{figure}

Figure~\ref{fig:flowchart} illustrates the working principle behind the adaptive e-ARQ procedure. Initially the wider beams are utilized, as there is no information regarding the target's position. Since a target is present, the reflected signal echo is received at the Rx node, with one of the available Tx/Rx beam configurations. 
The detection hypothesis, i.e., whether a target is present, is directly tied to the RESI, as a target's presence and well-matched sensing configurations shall lead to a strong received echo. As previously discussed, two different thresholds for this test parameter highlight the compatibility of the current configurations for the sensing task. When the RESI drops below the ACK threshold, but remains above the NACK threshold, a beam sweep procedure is triggered and a better suited configuration can be found.

\begin{figure*}
    \centering
    \includegraphics[width=0.95\textwidth]{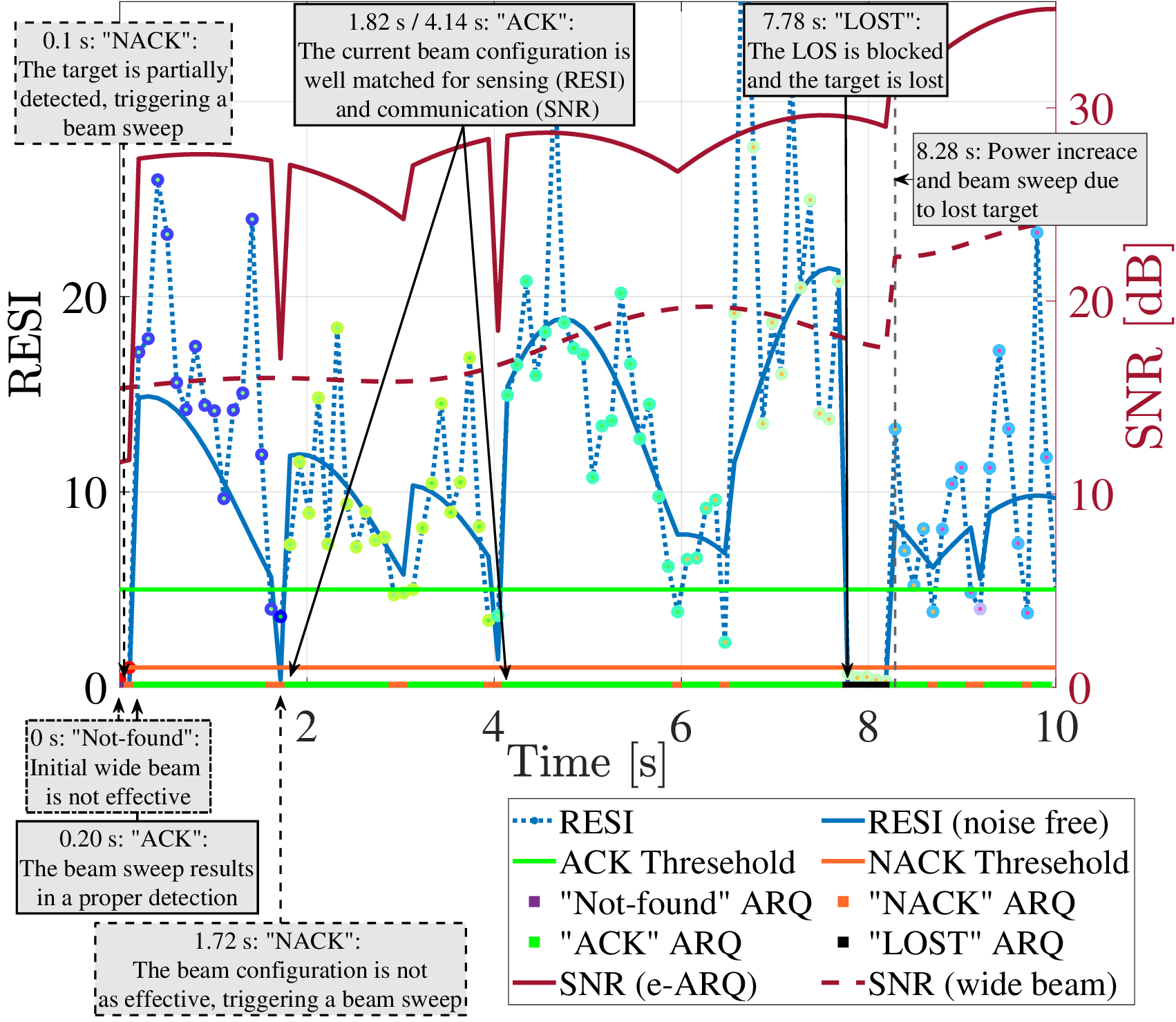}
    \caption{RESI and SNR over time and the corresponding ARQ for different Tx/Rx configurations. The target and the UE are assumed to be co-located, to simplify the discussions. The colours of the RESI marker ring and center indicate the sensing configuration utilized by the Rx and Tx node, respectively, for the corresponding iteration. The colours on the time axis indicate the corresponding ARQ response.}
    \label{fig:resi_snr}
\end{figure*}

Figure~\ref{fig:resi_snr} shows the RESI and the e-ARQ responses, triggered over time, when a small obstacle (see Table~\ref{tab:sim_parameters}) is placed in front of the Rx node. Initially, the target is not found and there is no indication that a target may be present, leading to a Not-found feedback. When the RESI rises above the NACK threshold the e-ARQ procedure dictates a possible target and there is a beam sweep procedure that results in proper detection (observe the RESI up to 0.20 s). Note that colours of the RESI marker ring and center in Fig.~\ref{fig:resi_snr} indicate the sensing beams utilized by the Rx and Tx nodes, respectively, for the corresponding iteration. As the target moves, the Tx/Rx node configuration remains unchanged for some time without an issue and the beam sweep procedure is not requested (see, e.g., time 0.20 s to 1.62 s). However, the compatibility of the current sensing configurations changes as the target moves, as indicated by the expected, noise-free, RESI. Eventually, the target moves outside the angular domain of the current Rx configuration and the RESI drops below the ACK threshold. The NACK ARQ triggers another beam sweep and the target is properly sensed with a better matched Rx beam configuration (observe the RESI between 1.72 s and 1.82 s). 

At some point of the target's trajectory (7.78 s), the LOS between the target and the Rx node is blocked by the obstacle and the RESI drops significantly. This results in a LOST feedback and a counter of 0.5 seconds is put into place. Since the target remains LOST, a drastic change is suggested and the transmission power is doubled (8.28 s), following the same trend as in typical ARQ procedures, e.g.,~\cite{makki2013green}. The counter ensures that the power is increased only when the target is truly lost. The additional transmitted power allows the Rx node to properly sense the target via the wall's reflection. As the target moves, different Tx beams are still required, despite the increased transmission power, due to the target's movement. The corresponding NACK feedback initiates the necessary adjustments.

In Fig.~\ref{fig:resi_snr}, the SNR at the UE inside the vehicle is also shown over time. It is shown that the initial Tx configuration provides poor communication quality. When our proposed e-ARQ procedure properly detects the target, the Tx node essentially creates a beamforming scheme that follows the UE's movement. Thus, the SNR remains stable and a reliable communication link is maintained. Without the e-ARQ procedure and the adaptive beamforming, a wide-beam is used, covering a large angular domain. On the other hand, the adaptive beamforming allows for properly designed narrow beams and, thereby, higher received SNR. Note that when the sensing application is disrupted (NACK ARQ) and a beam-sweep is necessary, the communication procedure is also affected, since it leads to improper Tx configurations. Once the UE is found (ACK ARQ), proper Tx configuration is implemented and the SNR is restored (see, e.g., time 4.14 s in Fig.~\ref{fig:resi_snr}).

The different RESI thresholds and the automatic procedure that each feedback corresponds to, shape the behaviour of the system and its performance. Table~\ref{tab:ARQ_table} demonstrates the probability of different triggered e-ARQ based feedback for a target moving in a random straight trajectory. 
The non-line-of-sight (N-LoS) probability in Table~\ref{tab:ARQ_table} refers to the obstacle preventing a direct connection between the Rx node and the target, while the request for additional resources marks that the target has been lost at least once during the simulation. Here, considering the simulation setup of Fig.~\ref{fig:geometry}, the results are presented for different sizes of the obstacle as defined in Table~\ref{tab:sim_parameters}.

It can be seen that different feedback are triggered with different probabilities. Importantly, as demonstrated in Table~\ref{tab:ARQ_table}, depending on the target moving trajectory and the obstacle size, different types of feedback and their corresponding adaptations may be required. The high percentage of ACK responses, relative to the NACK responses, indicates that our proposed adaptive sensing procedure successfully avoids a large number of beam sweeps. In other words, the sensing configurations are adjusted \emph{only when necessary}. This, not only improves the sensing quality, but also avoids sensing-to-communication interference and opens up resources for communication. Moreover, it can be seen that additional resources are commonly necessary for a reliable tracking procedure, highlighting the significance of the LOST feedback. The adaptive e-ARQ procedure utilizes the LOST response to effectively detect when a target's trajectory requires more power for the multipath echo to be detected and, thus, avoids unnecessary power consumption. The considered geometry and the required sensing characteristics need to be taken into account when designing the automated procedure tied to each feedback. 

\begin{table}
\caption{Percentage of different e-ARQ signals generated for different conditions (different sized obstacle).}
\label{tab:ARQ_table}
\begin{center}
\begin{tabular}{ |c|c|c| }
 \hline
                & Small Obstacle & Large Obstacle \\ 
\hline\hline
ACK$^{\ast}$                                & 84.22 \%       & 77.89 \%       \\ 
NACK$^{\ast}$                               & 10.51 \%       & 13.94 \%       \\  
LOST$^{\ast}$                               & 2.88 \%        & 3.88 \%        \\  
Not-Found$^{\ast}$                          & 2.40 \%        & 4.29 \%        \\   
N-LoS$^{\ast}$                              & 2.53 \%        & 31.23 \%       \\  
Additional resources requested$^{\ast\ast}$ & 36.40 \%       & 60.30 \%       \\ 
\hline
\multicolumn{3}{|c|}{$^{\ast}$Percentage per sensing iteration.} \\
\hline
\multicolumn{3}{|c|}{$^{\ast\ast}$At any point within a target's trajectory.} \\
\hline
\end{tabular}
\end{center}
\end{table}

Finally, with respect to our proposed e-ARQ scheme, the following points are interesting to mention: 

\begin{itemize}
    \item While we presented the setup for the ISAC networks, the same approach is well applicable to sensing networks, where the handshaking between the Tx node and the Rx sensing node can improve the sensing quality.
    \item Our proposed scheme is well applicable to complex multi-static ISAC networks with multiple Tx and/or RX nodes or the cases where the targets and the UEs are not co-located. 
    \item The presence of the e-ARQ based feedback can well affect the resource allocation trade-offs in ISAC networks. For instance, the sensing reconfigurations during the retransmissions -adapted to better match the target's position- would affect the communication performance, which would introduce, e.g., beamforming restrictions for the sensing/communication beams and/or interference management issues. Moreover, increasing the power allocated for the sensing application would reduce the power transmitted for communication purposes. This opens up different research questions to be answered in e-ARQ assisted ISAC networks.
    \item To enable e-ARQ, new signaling procedures needs to be defined in the standardization; Considering  wireless communications, ARQ is a well-established procedure. Particularly, in 5G New Radio (NR), 3GPP has already specified various signaling procedures for ACK/NACK responses in both downlink and uplink. Here, the specific procedures depend on the context of the communication, such as data transmission or control signaling. Although the standardization work on ISAC has not yet started by 3GGP, one may follow similar procedure as in communication-based ARQ, to define the appropriate signaling for e-ARQ covering sensing functionalities. Here, one may add indication flags to distinguish between the communication- and sensing-based feedback. Moreover, because e-ARQ is based on four different levels of feedback, as opposed to typical ARQ having only ACK/NACKs, more resources are required for the e-ARQ based feedback.
\end{itemize}


\section{Conclusions}
We introduced the concept of e-ARQ for sensing purposes. As illustrated, depending on the sensing status, the Rx node may provide the Tx node with different levels of feedback. Also, depending on the sensing status and the received feedback, the transmit and the sensing configurations may be adapted at the Tx and Rx node, respectively, during the retransmissions. Such a handshaking between the nodes results in accurate sensing, an improved understanding of the environment and, in turn, better spatial awareness. As a result, our proposed feedback procedure may affect the appropriate resource allocation between the communication and sensing functionalities, which can open up various research problems in ISAC networks. Finally, the design of proper sensing schemes that combine the sensing results in different retransmission rounds with our proposed e-ARQ signaling scheme, is an interesting research topic.

\bibliographystyle{ieeetr}
\bibliography{references}

\newpage

\begin{IEEEbiographynophoto}{Georgios Mylonopoulos}
is with the Department of Electrical and Information Engineering, University of Cassino and Southern Latium, 03043 Cassino, Italy, and also with the Consorzio Nazionale Interuniversitario per le Telecomunicazioni (CNIT), 43124 Parma, Italy and also with Ericsson Research, Ericsson, 417 56 Göteborg, Sweden (e-mail: georgios.mylonopoulos@unicas.it). His research interests include localization, ISAC as well as reconfigurable intelligent surfaces.
\end{IEEEbiographynophoto}
\begin{IEEEbiographynophoto}{Behrooz Makki}
(Senior Member, IEEE) is with Ericsson Research, Ericsson, 417 56 Göteborg, Sweden (e-mail: behrooz.makki@ericsson.com). His research interests include ISAC, MIMO as well as wireless backhaul.
\end{IEEEbiographynophoto}
\begin{IEEEbiographynophoto}{Stefano Buzzi}
(Senior Member, IEEE) is with the Department of Electrical and Information Engineering, University of Cassino and Southern Latium, 03043 Cassino, Italy, and also with the Consorzio Nazionale Interuniversitario per le Telecomunicazioni (CNIT), 43124 Parma, Italy. He is also affiliated with Politecnico di Milano, Milan, Italy (e-mail: buzzi@unicas.it).  His research and study interests are in the wide area of statistical signal processing and resource allocation, with emphasis on wireless communications and radar applications.
\end{IEEEbiographynophoto}
\begin{IEEEbiographynophoto}{Gábor Fodor}
(Senior Member, IEEE) is with Ericsson Research, Ericsson, 16480 Stockholm, Sweden, and also with the School of Electrical Engineering and Computer Science, KTH Royal Institute of Technology, 11428 Stockholm, Sweden (e-mail: gabor.fodor@ericsson.com). His research interests include signal processing and optimization for wireless communications
\end{IEEEbiographynophoto}

\end{document}